\documentclass[journal]{IEEEtran}
\usepackage[noadjust]{cite}

\ifCLASSINFOpdf
\else
\fi
\usepackage{amsmath}
\usepackage{todonotes}

\usepackage{stfloats}
\usepackage{url}
\begin{document}
\title{Encouraging early mastery of computational concepts through play}

\author{H. M. Dee,
        J. Freixenet,
        X. Cufi,
        E. Muntaner Perich, V. Poggioni, M. Marian, A. Milani% <-this % stops a space
\thanks{H. Dee is with the Dept. of Computer Science, Aberystwyth University, UK}% <-this % stops a space
\thanks{J. Freixenet, X. Cufi and E. Muntaner Perich are with Udigital.edu, Universitat de Girona, Spain}% <-this % stops a space
\thanks{V. Poggioni is with the Dept. of Mathematics and Computer Science
University of Perugia, Italy}% <-this % stops a space
\thanks{M. Marian is with the Dept. of Computers and Information Technology, University of Craiova, Romania}%
\thanks{A. Milani is with the Dept. of Mathematics and Computer Science
University of Perugia, Italy}}


\markboth{ArXiv Preprint}%
{Dee \MakeLowercase{\textit{et al.}}: Encouraging early mastery of computational concepts through play}



\maketitle

\begin{abstract}
Learning to code, and more broadly, learning about computer science is a
growing field of activity and research.  Under the label of \emph{computational
thinking}, computational concepts are increasingly used as cognitive tools in
many subject areas, beyond  computer science. Using playful approaches and
gamification to motivate educational activities, and to encourage exploratory
learning is not a new idea since play has been involved in the learning of
computational concepts by children from the very start. There is a tension
however, between learning activities and opportunities that are
completely open and playful, and learning activities that
are structured enough to be easily replicable among contexts, countries and
classrooms.  This paper describes the conception, refinement, design and
evaluation of a set of playful computational activities for classrooms or code
clubs, that balance the benefits of playfulness with sufficient
rigor and structure to enable robust replication.
\end{abstract}


% \fi
\IEEEpeerreviewmaketitle

\section{Introduction}

This paper describes the development of a set of shared, ready-to-use
activities that can be run in schools by teachers, in code-clubs by volunteers,
or by university staff in schools as \emph{outreach} activities.  These
activities are designed to promote and engage children with Computational
Thinking~\cite{wing,wing2,brady,peteranetz,compthinking2018}, in a playful and open-ended way.
All the activities have been tested and implemented by several teachers and/or
academics, in several countries, and the workshop material is organized in
order to emphasize opportunities to re-use and encourage context specific
adaptation.  

This paper opens with an analysis of the playful approach \cite{SEABORN2015} to
coding concepts, motivated by research into play and taxonomies of play. The
content and thematic arrangement of the proposed activities is then described.
Finally, the process used to generate and manage the workshop activities is
presented by detailing the phases of design, refinement, test and editorial
control.  An earlier version of this work appeared in~\cite{iticse17}: the work
presented here has been extended to integrate taxonomies of play and
playfulness within the proposed workshop activities, and to further explicitly
clarify the \emph{computational thinking} aspects of the workshops. The contributions of 
of this paper are:

\begin{itemize}
\item \textbf{To explicitly involve play in the learning process} through a clear analysis of taxonomies of play, and their relationship towards various computational thinking concepts and learning activities;
\item \textbf{To share the process} through which playful workshops that engage school pupils in computational thinking can be iteratively refined and edited in order to maximize their reuse potential;
\item \textbf{To show computational thinking concepts embedded in a set of cross-curricular activities};
\item \textbf{To encourage reuse and contextual adaptation of materials} through offering tested-in-the-field workshops to a broader computing education community.
\end{itemize}

Often a lot of excellent, playful and fun \emph{outreach} work is trapped
inside institutions. The proposed framework for activity representation,
sharing and improvement presented here aims to shed light on these quality
experiences.  Through this we encourage colleagues to grow creatively and to
engage with resources that will involve K-12 students in computing at all
levels. A more long term goal is to encourage creative and passionate teachers
to use this methodology for proposing, sharing and passing on  their own
original and creative contributions.

\begin{figure}
\begin{center}
\includegraphics[width=8.5cm]{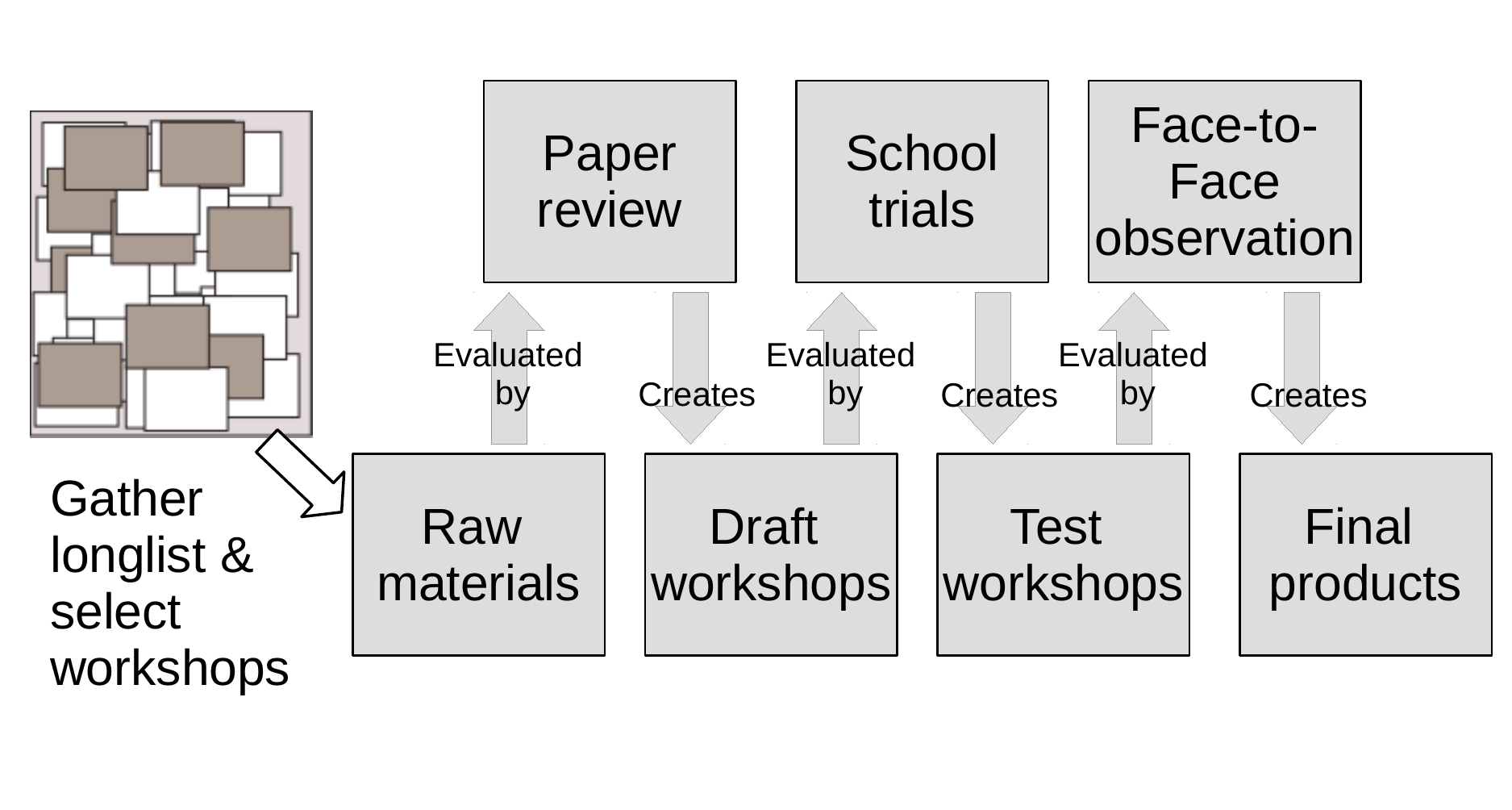}
\end{center}
\caption{An overview of the workshop creation process\label{f:process}}
\end{figure}

The proposed methodology for creating re-usable materials and workshops is
shown in Figure~\ref{f:process}. The main process can be summarized as:

\begin{enumerate}
\item Gather candidate playful workshop proposals;
\item Select a subset of workshops for further development according to our criteria;
\item Write selected candidates in draft form using a common structure providing \emph{raw materials};
\item Perform a preliminary paper-based review;
\item Revise to obtain \emph{draft workshops};
\item Test the workshops in schools, independently;
\item Revise them based upon school experience feedback creating thus \emph{test workshops};
\item Test them in closely observed conditions; 
\item Finally revise, and release the workshops material to the web.
\end{enumerate}

The proposed process emphasizes quality since it produces a workshop that has
been thoroughly tested and refined in several situations -- in the case of the
project experience described here, several schools across several countries.
In the following section, the local context for each of the participating countries is
described briefly.
 
\section{The local context}
\label{s:context}

The methodology and the experience here described is part of the outcome of a
large-scale European Union project\footnote{Early Mastery project -
http://playfulcoding.udg.edu/about/} involving participants from 5 countries,
with four universities, two schools and a start-up company as core members.
Workshops and activities have been tested across several additional contexts
and events, including university visits, code clubs, science fairs and over 80
different schools.  The diversity of the project participants (all European,
but from different educational backgrounds, educational funding landscape,
computational context, and regulatory systems) helped ensure a wide
applicability of the resulting activities. In particular, the relationship
between computer science and the school curriculum was not uniform across
project teams, requiring a flexibility about computational context, classroom
resources and realized experience. If the same activity works well in urban
Spain, small-town Romania, and rural Wales we believe it has very broad appeal. 

Across Europe, computing education has been changing, 
and the introduction of more computational thinking, practical
computer science, coding and ``informatics-type'' activities is becoming
a widespread curricular move~\cite{fref1,learningwales,sdar,cata,pnsd_Italy,pf_Italy,eu16,rref1}. 
The transition, frequently observed,
ranges from educational systems in which computing enters the 
classroom thanks to a motivated teacher to organizations in which 
computing is a mandatory part of the curriculum~\cite{eu16}.
This transition is rarely smooth.
New subjects in the curriculum require new lesson plans, 
new teacher training, new approaches to learning and  
often new hardware or other types of equipment. In this context, university outreach 
efforts can become key sources
of additional support for schools and colleges trying to keep up with
the pace of curricular change.

In summary, IT was well-embedded across the curriculum for many of the
project's participant countries. It is not unusual to see spreadsheets in
History class, or word processing in English.  Computer science and
programming, however, were rarely seen outside computing classes.  In this
project, innovative ways have been proposed to foster creative and critical
learning across the curriculum through the playful use of programming and
robotics. 

\section{Why playful?}
\label{s:playful}
Our commitment to providing creative and playful activities is based 
upon several ideas. First, there is a common belief that playing and games \cite{SEABORN2015}, particularly in the early stages
of learning computing, can be a driver for progress. Exploration and
self-directed learning are key to constructivist ideas~\cite{papert1980}, and
the concept of playing as a ``leading activity'' comes earlier, from Vygotsky~\cite{vygotsky67}.  In this formulation, play represents a social construction that
allows a child to move beyond their current ideas, developing new mental
processes, \emph{leading} to cognitive development.

This exploratory, playful approach has been part of learning to code for
children since Papert's work on Logo~\cite{papert1980}. Moving from exploration
and open-ended discovery~\cite{burlson}, to playful exploration and to active use of play in
learning to code is one of the aims of the current work. It is important to avoid the suggestion that digital play is always educational: some educational software is playful but not open-ended, educational claims are often linked to
marketing, and interactivity sometimes provides initial motivation but no
further depth. As Stephen and Plowman state in~\cite{stephen14}, ``\emph{Digital
interactivity alone does not guarantee either an educational or a playful
encounter}''.  

These concerns lead us to our second key idea: it is useful to think more
deeply about what it is meant by ``playful'' by further considering taxonomies
of play, and the various ways in which play has been categorized by researchers
in education and in child development~\cite{hughes2002,else2014}. Considering
the variety of different play types can greatly assist planning and structuring
when creating playful coding experiences. Play categorizations and taxonomies
include hierarchical systems, as are applied in~\cite{birdplay14} (where
\emph{epistemic} play is distinguished from \emph{ludic} play, and then further
subdivided), or observational.  Whilst these have generally been developed with
outdoor, physical play in mind, the applicability of these to the digital world
is clear (as emphasized in~\cite{marsh2016}). These taxonomies of play types
derive from extended observations of children playing in the real world or in
virtual worlds, and break down the activities children engage in whilst playing
into specific categories.  

All the categories from~\cite{hughes2002} are applicable to the digital world
in some sense~\cite{marsh2016}, but in this work, a more concise taxonomy and a
selection of subset of play types is proposed. The main reason for the
restriction is that original general play categories are less applicable to
computer-based activities (\emph{rough and tumble play} and \emph{locomotor
play} for example rely on physical activity).  The subset of play type for
digital activities is presented in Table \ref{t:playtypes} and is confined to
those types of play which can be found in the classroom within a computational
learning context although others may occur in ``CS-Unplugged'' type
activities~\cite{Nishida2009, tangney}.  The list shows these play types in the
approximate order of their popularity in a computing education context. Play is
ubiquitous in computing, indeed ``\emph{Have a play with it}'', meaning ``try
it out and see if you can work out what it does'' is the first step for most
expert users of technology presented with a new tool. This activity maps
exactly on to what Hughes describes as \emph{exploratory play}.

\begin{table*}
{\def\arraystretch{1.4} %makes the table cells have vertical space
\begin{tabular}{p{2.8cm}  p{14.8cm}}
\hline
\textbf{Play type} & \textbf{Brief description} \\
\hline
Exploratory & Using the senses to explore and discover possibilities or find out information \\
Mastery & Play in which the players try to gain control of a skill or environment, maybe through practice \\
Symbolic & Using one object or item to stand in for another \\
Creative & Play that enables the player to develop ideas and make things \\
Communication & Using songs, rhymes, words and poetry in play \\
Dramatic & Dramatising events that the player has not been directly involved in \\
Imaginative & Play in which players pretend that things are otherwise \\
Object & The manipulation of objects and things through play \\
Role play & Play involving the adoption of different roles \\
Fantasy & Taking on roles which could not occur in real life e.g. superhero \\
Transgressive & Play which involves the player pushing at boundaries, for example, bending the rules of the game \\
\hline
\end{tabular}
}
\caption{\label{t:playtypes}A subset of play taxonomies which can be used to conceptualize digital activity, adapted from~\cite{hughes2002,marsh2016}}
\end{table*}

Through a consideration of these play taxonomies, is possible to understand the progress in workshops in terms of types of play. Thus an important consideration~\cite{mclaren2008} becomes whether children are able to move 
from exploratory play (asking ``\emph{what does this do?}'') 
to imaginative or creative play (asking ``\emph{what can I do with this?}'', 
or ``\emph{what can I make this do?}''). 
The category of play least likely to be
found in the classroom context is that of transgressive play, or play which is 
crossing and stretching boundaries. The closest approximation of transgressive play that can be found in the proposed workshops,
comes from robotics activities; for example, with some hardware configurations it is possible to make a robot ``do a wheelie'' by reversing direction rapidly. Once children discover this, they sometimes modify their code to deliberately cause and repeat the effect.

Thirdly, and perhaps most importantly, these workshops are designed to
teach computational concepts, and also to encourage students (along with their 
teachers and parents) to see that computing can be
\emph{fun}. 
In a world in which many countries still do not have a formal computing
curriculum, and in which others have a very formal and mathematical-oriented curriculum~\cite{aleven, prieto}, 
workshops emphasizing the creative and playful nature of computational skills have a key role to play, particularly in attracting a more diverse range of future computer scientists. We are not only computer science educators, we are also computer science evangelists.  

Storytelling is something of a special case, and major potential as a workshop
motivator. It brings together a particular subset of play types (creative,
imaginative, communication, dramatic) and also involves the construction of an
artifact (a story)~\cite{harari2014}. But on top of that, there are a number of
more general reasons for encouraging storytelling in the classroom:

\begin{itemize}
\item It increases the enthusiasm to read and re-read, because doing so allows to discover stories;
\item It increases the motivation to write, as a means to express stories~\cite{kori};
\item It improves \emph{soft skills}: the ability to listen, and to express ourselves publicly;
\item Creating and telling stories allows participants to project and express their own emotions, feelings, and thoughts;
\item Sharing stories helps participants put themselves in the role of others, developing empathy (and engaging in \emph{role play};
\end{itemize}

Digital storytelling is a relatively new term that refers to stories that
include multimedia elements such as photographs, videos, sounds, texts, and
also narrative voices, and this has found its way into the classroom in a
number of contexts~\cite{Smeda2014}. Robin in~\cite{robin08} provides a framework for thinking
about digital storytelling in the classroom; authors have shown that digital
storytelling can help visual memory~\cite{sa:1}, creativity~\cite{sp:1,gresham2014} and it
can also improve academic achievement~\cite{ya:1}. Creative computing,
programming, or robotics fit this framework well. It is possible to program digital
stories and animations with virtual characters and scenarios, but physical characters and scenarios can also be programmed in the form of robots or
constructions outside the computer. Creative computing adds a new component to
stories: \emph{interactivity}.  

\section{Workshop Selection}
\label{s:content_overview}

The core of this project was the creation of a set of resources 
which could be used, re-used and modified by 
university staff engaged in outreach and collaborating with school teachers 
in search of innovative lesson plans. The starting point for this
was a long-list of ideas and half-tested activities, written by
the project partners.  
This long-list was cut down to a set of around 20, based
upon inclusion criteria directly related to the main objectives 
of the project. All included workshops were addressing in some way, the following points:

\begin{itemize}
\item \textbf{Explore new ways to promote teaching and learning of computer programming in European schools}; 
\item \textbf{Help school children move from being digital consumers to digital creators};
\item \textbf{Make it easy for people to share the results of the project};
\item \textbf{Inspire schools to use computer programming in an interdisciplinary manner};
\end{itemize}

In addition, the workshops were chosen for their \emph{playful} nature: in fact,
the project was nicknamed ``Playful Coding'', and the use of play within
the activities was key to their inclusion. Thus, less emphasis was given to workshops
with explicit, clear step-by-step instructions, and more interest and effort has been spent in 
workshops that allowed the participants time to experiment (exploratory play), 
repeat (mastery play), create and imagine (creative and imaginative play).

The second objective is explicitly linked to ``maker'' culture,
constructivism~\cite{papert1980} and computational creativity.  The move from
consumer to creator is one that many organizations are now championing, e.g.
the World Economic Forum~\cite{wef12}.  Workshops that lead to products or
stories, such as animations, movies, games or applications have this enabling
aspect. This emphasis on creativity and product implies that workshops were
more likely selected if they would involve active learning
methodologies~\cite{crc2006}, inviting students to create, design, modify and
share. In these activities, technology is not an end in itself but merely a
means to express creativity.

With regard to the third objective, there are some aspects of the global
project that are vital in terms of sharing.  The outputs are made accessible to
the world, via a web platform~\cite{playfulweb}.  Although the language of the
platform is English, the Teachers' Guide~\cite{playfulbook} and complementary
material have been translated into the languages of the project (English,
Spanish, Romanian, French, Italian and Welsh).  With regard to workshop
selection, priority goes to workshops with minimal
setup cost, free software, and open/platform neutral environments. The project
includes participants using Apple, Microsoft and Linux environments; it is
important to be able to support all platforms for maximum school applicability.
Through encouraging reuse and contextual adaptation of the provided materials,
teachers are encouraged to see how computing can inspire activities across the
curriculum, leading themselves to develop further resources.

The fourth objective led to select examples of introduction of computer
programming and robotics into school curricula in the broadest sense. Workshops
in the final selection include a wide range of subjects including languages,
poetry, geography, art, and many others across the curriculum. 

\section{Refinement of workshops}
\label{s:process}

An extensive testing and refinement process (illustrated in
Fig.~\ref{f:process}) has been applied to the selected activities. This process
consists of three sub-phases, resulting in an iterative improvement of the
activities in terms of usability, structure, and presentation impact.  Each
partner recorded a summary of the workshops that they had proposed, providing
the \emph{raw materials} (initial workshop proposals). The \emph{raw materials}
were preliminarily subjected to a paper-based review by a project partner based
in a different country (cultural context) to ensure cross-cultural
applicability.  Workshops were then revised based upon feedback to
generate a \emph{draft workshop} ready to become an in-class experiment.
In the second phase, these \emph{draft workshops} were evaluated in school
trials and iteratively improved again, generating the \emph{test workshops}.
\emph{Test workshop} proposals, already evaluated by several partners, were
implemented in different countries with face-to-face observations, and reassessed
in order to generate the \emph{final products} in the last
phase~\cite{strijbos}.

\subsection{Paper review}

Workshops were initially developed by each partner in close collaboration with
their local community, including teachers, schools and after-school clubs.  It
has been observed that the first version of a workshop often carried the
imprint of the local setting and the influence of national educational context.
It has been also found that when designing playful and creative activities the
input of several people can help build a richer experience. As an example, the
first iteration of a workshop might involve some coding in Scratch
\cite{Resnick09} to make a quiz, and suggestions from the paper review
might add asset creation (drawings, photos), swapping code to play
with and test one another's work, and other game-like elements such as timers
and high-score tables.  Feedback was then integrated to generated the
\emph{draft workshop} revisions for actual in-the-field testing at a 
school. 

\subsection{Distributed testing: school trials}

\begin{table*}[!ht]
{\def\arraystretch{1.2} %makes the table cells have vertical space
\begin{tabular}{p{11cm}  p{6.3cm}}
\hline
\textbf{Question} & \textbf{Answer} \\
\hline
How much time did it take you to prepare for the activity? & [{\it Numeric answer}] \\
Please give a global mark for how easy it was to understand the activity before implementing it & [{\it 1-5}]\\
How much time did it take you to implement the activity with kids? & [{\it Numeric answer}]\\ 
Please give a global mark for how easy it was to implement the activity & [{\it 1-5}]\\
Do you think the proposed age range is adequate? & [{\it Yes/No}] [{\it Comments}]\\
Would you recommend this coding activity to other teachers/schools? & [{\it Yes/No}] [{\it Comments}]\\
Is the goal of the activity clear enough? & [{\it Yes/No}] [{\it Comments}]\\
Did you achieve the goal of the activity? & [{\it Yes/No}] [{\it Comments}]\\
Did the activity reach your expectations? & [{\it Yes/No}] [{\it Comments}]\\
Did kids enjoy the activity? & [{\it Yes/No}] [{\it Comments}]\\
Do you think kids have developed new skills while working on this activity? & [{\it Yes/No}] [{\it Comments}]\\
If yes, please, indicate which skills have they developed: &
[checklist including {\it Heuristic method (Trial-Error);  Communication Skills;  Computational Thinking; Creative thinking;  Problem solving } \ldots]\\
What do you think kids like the most/least? & [{\it Free text}]\\
What would you change in terms of description of the activity? & [{\it Free text}]\\
Please give a final, overall mark for the activity & [{\it 1-5}]\\
\hline
\end{tabular}
}
\caption{A summary of the teacher feedback form from the distributed testing phase. These questions were preceded by a series of questions about the school and the children in the classroom (age range, class size, public/private school, and so on)} \label{teacherform}
\end{table*}

The next stage of activity refinement involved testing by the partners within
the project consortium.  Each partner was involved in organizing workshop
trials with classes of local schools. It has been ensured, as much as possible,
that each workshop was tested with classroom groups in at least two additional
countries, beside the originating country. This allowed the project to provide workshop
authors with feedback from outside the original social, national, economic and
computational context. During trial sessions, the classroom teachers were
asked to take notes, and upon completion of the trials, they were asked to fill
in a simple web feedback questionnaire about the workshop. Only two workshops
were difficult to test outside their home environment. These
were both robotics workshops and the reason for difficulty was
equipment availability in testing sites. The content of these
workshops was approximated using alternative wheeled robots and modified 
accordingly.

The questions of the web teacher feedback questionnaire are shown in
Figure~\ref{teacherform}. As many other aspects of the project, the teacher
feedback questionnaire underwent through several refinement iterations before
stabilizing in the presented form.  Some teachers are very keen to assist with
the project, and will fill in a lot of detail. But other teachers, under
strong time pressure, are less keen to provide detailed textual
answers~\cite{prieto}. Therefore, the form evolved into a structure with
predominantly yes/no questions, augmented with  text boxes to enable the more
willing teachers to fully contribute.  Teachers were asked to complete the
feedback form immediately, or soon after the workshop had been completed, and
data was entered into the repository directly by means of Google Forms.

In earlier iterations, feedback was also collected from the children
participating in the workshops; this proved very positive and confirmed  the
workshops were playful, interesting and fun for the children.  However, the
children's feedback was not as directly useful as teachers' feedback for further
workshop improvements due to a lack of critical/negative comments. 
%For example when asked ``\emph{what age would this activity be OK for?}'' all
%children said something equivalent to ``My age and up''.  \todo[inline]{think
%about this previous paragraph and tidy up what we're saying}

\begin{figure}
\begin{centering}
\includegraphics[width=8cm]{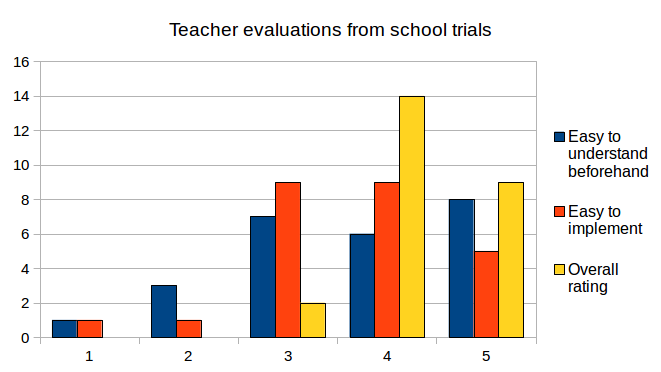}
\caption{\label{f:teacher_eval} Teacher judgements of ease of comprehension before implementation, ease of implementation, and overall rating after the workshops.}
\end{centering}
\end{figure}

Investigating the aggregate feedback from these school trials indicates that
many teachers had some preliminary concerns when they first encounter the
activity. Teachers' ratings (on a 1-5 scale) capture how easy they believe the
activity is going to be to implement, how easy it was to implement, and an
overall rating for the activity.  Overall ratings were always higher, than
those for representing teachers' perceptions of the workshop beforehand.  This
is probably not surprising, considering that the teachers involved in the
project were often non-specialists. The chart in Figure~\ref{f:teacher_eval}
shows clearly this change of rating. All except one of the responses to the
question ``Would you recommend this coding activity to other
teachers/schools?'' were positive. The one negative response was associated
with a workshop teaching Arduino and C programming for robotic control, which
was thought to be too advanced for the class participants.

Feedback on specific workshops from these distributed school trials 
was passed back to
the workshop originator, and the workshop description was then 
updated as required. 

\subsection{Face-to-face workshop observation}

\begin{figure*}
\begin{tabular}{| p{0.95\textwidth} |}
\hline
\textbf{TASK}: To observe and detect those aspects you consider important to take into account for the evaluation of the activity. Please take as a guideline the following items for the analysis: \\
\textbf{Methodological Aspects}: Introduction, presentation of the activity, rhythm of the work, timing, types of intervention during the activity, types of groups (individual, in pairs, in groups \ldots), time spent on the activity, reorientation of errors.  \\
\textbf{Interaction types}: teacher-student, student-student, student-material. \\
\textbf{Learning process}: playful, trial and error, planned strategy, collaborative work. \\
\textbf{Evaluation}: follow-up work, collection of evidence and results, observation guidelines. \\
\textbf{Classroom environment}: number of students, materials, space distribution, noise, orderliness, stimulus. \\
\hline
\end{tabular}
\caption{The classroom observation criteria for face-to-face testing sessions\label{colleagueobs}}
\end{figure*}

The final stage of workshop revision involved dedicated 
training events in which educators from different schools 
were asked to perform the activities with classes of pupils. 
This took place during several intense weeks of workshop evaluation 
in which
the project as a whole visited schools in the UK and Spain, and hosted
school visits in France. These sessions were observed by 
project partners from  different institutions, taking notes 
on various aspects of
the activity. Verbal feedback was also solicited, from 
the observers and the observed instructor, in a round-table feedback
session immediately following each workshop.  
Figure~\ref{colleagueobs} shows the observation criteria
for the written aspect of this face-to-face evaluation.

As before, the observers were oriented to look explicitly for creative,
collaborative work in a playful context, and for constructive workshop
improvements. This process resulted in several pages of handwritten
feedback for each workshop leader, with comments on all 
aspects of the learning experience from content type to classroom
management. The feedback received from these
sessions had much greater depth than that received from
the earlier distributed in-school trials. The feedback was then incorporated into 
the workshop description, to provide our final iteration
of improvements. 

%This was a costly phase, and we were fortunate enough
%to receive funding for travel, facilitating such intensive
%face-to-face evaluation.
%The feedback received from these
%sessions had much greater depth than that received from
%the earlier distributed in-school trials. This is perhaps because
%all of the participants were familiar with the aims of the
%project, but maybe also because they were familiar 
%with each other, enabling robust comments
%in a way that is otherwise difficult in a ``school visit''-only scenario.
%At the end of this phase, activities have 
%been read through, amended, tried in two new environments,
%amended, observed in a new setting, and finally modified once more.

\section{Content of the workshops}
\label{s:content}

The output of this lengthy refinement process has been a set of workshops that
have computational thinking at their heart, and which combine different types
of play in order to motivate and explore a variety of computational thinking
concepts. In this section, some the proposed workshops are described in detail,
and their possible concatenation and/or integration is presented. 

Most activities are based on the use of Scratch programming
language~\cite{scratch} in combination with other topics from the core
curriculum in order to amplify the interest of both pupils and school
teachers~\cite{Resnick09}.  This interdisciplinary connection could be seen as
taking advantage of the entertaining nature of Scratch for supporting  the
learning of specific topics from other subjects, but it could also be seen as
taking advantage of students' interest in other subjects to encourage
engagement with computing.  The observation is that both directions can benefit
from this situation: to quote Resnick, it is an opportunity to ``learn to code
and code to learn''~\cite{resnickted}.

The drive towards cross-curricular workshops led to an unconventional
organization of our materials. Rather than classifying workshops strictly in
one single category, ``menus'' of activities have been provided along
particular themes. As with a restaurant menu, it is possible for one dish to
appear in several different meals. In this way, it has been possible to
highlight the interdisciplinary nature of created activities whilst still
making the disciplinary connections explicit to potential users. Some examples
of the provided workshop thematic menus are shown in the next sections.

\subsection{Geography menu}

Three workshops explore various geographical concepts in a playful way.  The
first workshop uses basic Scratch ideas to allow participants to talk about
their country, town, or area. In this workshop, participants start with a map,
which they can draw themselves if time allows. This map is then decorated with
sprites (a typical Scratch entity) that represent parts of the area with
pictures for each city, or street, or building. Choosing images to use as
sprites that represent locations is a fun activity, and this can be made into a
group task depending upon time. Sprites are then animated so that clicking on
the sprites brings up information - for example, clicking on the sprite for
``Cardiff'' could bring up information about population, or sporting venues, or
parks\ldots

\begin{figure}
\centering
\includegraphics[width=8cm]{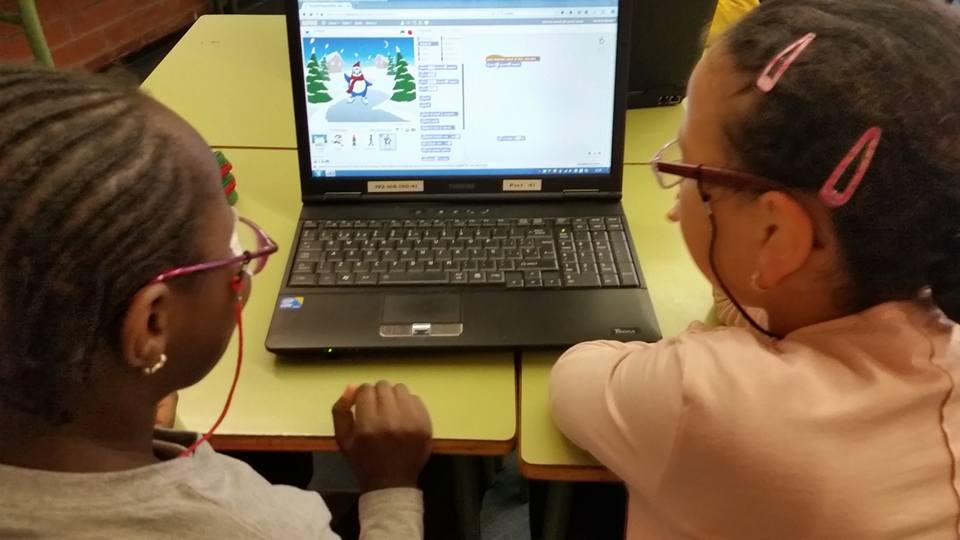}
\caption{Participants ``talking about the weather'' with Scratch\label{img_weather}}
\end{figure}

The second workshop is a coding activity with Scratch that aims to teach
vocabulary items, especially about weather and free time activities. Images
representing the four seasons are annotated with images for weather events, and
textual descriptions of these. This workshop originated as a language teaching
workshop for older students learning foreign language weather vocabulary, but
can usefully be used with younger students in the participants' main language.
Through this activity, participants learn about representations of weather, and
illustrate changes in the seasons.  This workshop is illustrated in
Figure~\ref{img_weather}.

The final workshop further challenges the children, by moving beyond Scratch
and introducing HTML. In this workshop, participants make a web page that can
load a dynamic weather forecast, provided on-line. This can lead to a
discussion about linking the world's geographical information through computing
in order to provide news about their local area/town/country.  For example, how
does their region compare to others in the world?  All of these three
activities combine aspects of playful interaction with creativity and design;
in all of them, participants are engaged with building something with software
(a Scratch program or a web page) that helps convey geographical facts that are
relevant to the students local area or town. 

\subsection{Mathematics menu}

The \emph{mathematics} thematic menu is specifically designed to motivate
students to explore mathematical concepts in a playful way~\cite{aleven,
augustin}. Mathematical concepts are often implicitly embedded in the work
rather than explicitly leading the activity.  Mathematics can be demotivating
for a subset of students, so activities aim to explore concepts like angles,
sizes, proportions and other geometric concepts through animation, illustration
and graphics.  Participants are asked to draw, program, count, play, cut paper,
measure, build things and animate. In a typical constructivist approach,
through these activities, the participants learn about angles, shapes and
geometry, while avoiding a potentially demotivating formal approach.

One example activity is the creation of a timer/chronometer with Scratch 
which has a dial with hands for seconds and minutes. 
The first step in this is to choose or create different programming “sprites”: 
a clock face, showing the seconds and minutes and two characters 
for the hands of the clock.  Once chosen, these programming objects are linked to
different sub-programs that will allow us to achieve other goals.
The mathematical challenge underlying this activity is  
to discover the angle that is necessary to turn the clock hand each second or 
minute. Advanced participants can go on to build a digital chronometer, or to
add hour hands, or ``pause'' and ``reset'' buttons. 

Another activity exploring mathematical concepts starts with a playful
photo-walk.  Participants walk around the building and the surroundings (or
just their classroom) with a camera, and take pictures of the regular
geometrical shapes that they see in their day-to-day life (rectangular windows,
circular traffic signals, etc.).  Pupils then collect their photos on a
computer with Scratch installed.  In teams, they build a program with Scratch
in which one of the images previously taken, forms a background. In the program,
a character (sprite) has to follow the contour of the geometrical shape
appearing in the picture. The same activity could be implemented with robots, for instance Lego Mindstorms, in which case the shapes the robot needs to
trace are drawn and marked with colored tape on cardboard or the floor.

\subsection{Storytelling menu}

As mentioned in Section~\ref{s:playful}, storytelling is a very important form
of communication, and one of the clearest ways to bring creativity and
playfulness into a computing context~\cite{slator}. Several storytelling-based
workshops have been created in the project; two examples of them are described
here.

Collaborative digital storytelling with Scratch involves the participants
working in teams to design and program a collaborative story.
Each team programs one part of a global story in their own computer, then they
synchronize the parts, and at the end of the activity all the computers are put
together in a row, and the students can watch the full story where the
characters move, speak and ``jump'' from one computer to the next.

At the beginning, the class decides on the topic and the main storyline. This
begins with brainstorming and storyboards with paper and pencils. Then, each
team works on creating the characters and backgrounds of their part of the
story (with traditional or digital techniques), and pupils can even record
their own voices for the dialogues.  They upload these creations to a new
Scratch project, and they start the programming the characters to move, speak
and perform the required actions. This is the part of the activity that
involves coding. For each section of the story, the characters should come in
from one side of the stage, and leave through the other, consistently (so the
story reads from left-to-right, for example).

After coding the entire story, the transitions between story sections need to
be synchronized. The easy way to achieve this is to work out how long each
animation section is, and then add an appropriate delay to the start of all
programs except the first one. When participants press ``go'' concurrently, the
program corresponding to a single story section will wait until its own turn.
Finally, the computers are positioned to form a line; in this way, all children
watch the final animation with the chained stories, and the characters will
appear to jump from one screen to the next. 

A very engaging further extension of this activity involves synchronizing the
stories by using physical sensors and motors instead of using timers. The idea
is to combine collaborative storytelling and chain reactions. One way to
implement this challenge is by means of LEGO WeDo sensors and motors, which are
compatible with Scratch. To make physical-world \emph{contraptions} some
physical-world materials are helpful: cardboard tubes, balls, tape, dominoes,
sticks, etc.  

Another example of story-related activity consists of creating an animated story from a
poem that has been previously studied in class, or a poem that children have
created.  The idea is to combine images, e.g. drawn by children or downloaded from
the Internet, with the verses of the poem. They have to appear sequentially, so
the final result is a visual version of the poem that could also include
recordings of their own voices reciting the verses. 
 
Children then use Scratch to program images and texts to appear and disappear
at the right time, following the rhythm of the language in the poem. This
involves timing, animation, and an understanding of the links between poetry
and imagery. This workshop again be adapted to the context of a foreign
language class, but can also be used in the context of literature classes of
the students main language. 

\subsection{Artificial intelligence}

Whilst coding is clearly a core competence for computational thinking,
computing is much more than just programming.  The AI workshop considers more
theoretical questions based around the Turing Test, and whether computers can
think. This workshop proceeds through a mixture of game-playing and mini
scientific experiments, discussing questions about what makes a thing
intelligent, and how one knows that a thing is intelligent. The games include:

\begin{itemize}
\item \textbf{Text-message Turing Test} - in which one of the pupils leaves the room with a helper, and the class sends questions by SMS to guess which person is answering;
\item \textbf{Intelligence ordering game} - in which a set of around 30 photographs of 
objects or organisms (a rock, a chess computer, a sheep, a kitten, a drone, 
a robot \ldots ) are ordered by perceived intelligence by the group. This is 
followed by a discussion of what qualities make a thing intelligent; is it 
language? Sight? Being able to make friends?;
\item \textbf{Can you trip up an AI?} - in which the pupils converse with
a chatbot, competing to get it to make the stupidest answer.
\end{itemize}

The mini science experiment involves coming up with three questions for
chatbots, and then trying these questions on different chatbots, recording the
answers, and trying to decide which chatbot is more convincing. The
participants are encouraged to think about what the question is testing, and
how they expect the chatbot to respond, forming hypotheses about the chatbot's
behaviour, and then testing these hypotheses.

\subsection{Robotics and hardware related workshops}

Robots are by their very nature motivating to some students~\cite{javier,
crenshaw, shim}. Students love seeing their program having an effect in the
real world: they can literally move things around with their ideas. However,
robots can be temperamental, and if they break, there can be catastrophic
effects for a workshop experience.  That said, the motivational aspects far
outweigh the risks. There is nothing more rewarding than seeing the sense of
achievement on a student's face when they have finally managed to get the robot
do what they wanted it to. 

Several workshops involving robots have been tested and refined through our
project, based upon three different hardware families (POB, Arduino-based
robots, and Lego Mindstorms).

In the case of robots, schools will need to use what they have, as the cost of
setup with new technology can be prohibitive. If schools have some other
platform (that is, some robot other than Arduino, POB or Lego robots), the key
ideas and concepts can be transferred to another wheeled robot. From a
computational thinking perspective, the robot workshops deal with ideas of
\textbf{control structures} (decisions, reacting to the environment),
\textbf{iteration} (looping), \textbf{debugging} and other development related
concepts, and perhaps most uniquely in child-focused workshops,
\textbf{precision} and imprecision.  Dealing with feedback loops, noisy sensors
and real-world robots is a great learning experience from a computational
perspective~\cite{seung}. 

\subsubsection{The general structure of a robot workshop}

All of the projects' robot workshops have a similar overall structure,
regardless of platform. This makes it easy to adapt the workshop structure for
use with any suitable wheeled robot. The basic steps are listed below:

\begin{itemize}
\item{\textbf{Move the robot} forward and backwards. Getting the robots moving early in the session is important for motivation, and will also expose any hardware faults at the outset;}
\item{\textbf{Make the robot follow a path corresponding to a shape} for example, a square, showing that the controls entered into the robot's program can make it move more-or-less precisely. This also involves combining two types of command (move and turn), or involves moving one wheel and not another (depending upon the dynamic properties of the robot);}
\item{\textbf{Modify your program so that the robot makes a different shape} for example a triangle, or a pentagon: this involves modification of the control code, and some calculation of angles.  Advanced students
at this point can be challenged to consider the problem of making a circular path;}
\item{\textbf{Make the robot repeat this} e.g. creating \emph{four} triangles;}
\item{\textbf{Make the robot read a sensor} and then react to the value of that sensor, through printing to a terminal, or making the robot do something (like beeping or flashing);}
\item{\textbf{Make the robot change behavior in response to a sensor} by avoiding an obstacle, or responding to a loud noise by changing the shape it draws.}
\end{itemize}

The details of these steps can depend on the platform~\cite{playfulweb}, but
taken together they provide the opportunity to learn more about loops, control,
precision, sensors, and debugging.

%\subsubsection{Tips for a successful robot outing}
%
%Robots, whilst motivating, can also be frustrating. We believe that the benefits
%outweigh the risks, and present here a set of tips for minimising the possibility of
%robot failure.
%
%\begin{itemize}
%\item \textbf{Power}. Make sure you have spare batteries, chargers, or whatever else you need to make the robots go.
%\item \textbf{Take Spares} if you can, take one more robot than you need in case of hardware failure.
%\item \textbf{Unreliablility} is one of the key lessons robots teach. Precision is hard. If you tell a robot to drive one meter, it will probably go 110cm. Or 90cm. If you are lucky, it will be consistent, but it might not be. Actuators and sensors suffer from noise. This is just the way robots are and it is one of the things we learn about through working with them.  This has to be treated as a \emph{feature}, not a 
%\emph{bug}.
%%\end{itemize}
%
%When choosing a robot for use in the classroom one size does not fit all.  A
%simplified programming environment can be very useful with younger pupils, but
%with older students, more traditional text-based programming (e.g. Arduino C) can
%be an excellent and challenging route.

\subsection{Packaging workshops for reuse}
\label{s:reuse}

The workshops represent the core of the project, and the process just described
has resulted in well-tested, standalone workshop packages and teaching material
for the use of educators outside of the project. The single workshops are
available on the project website~\cite{playfulweb}, as individual pages or as
part of thematic menus. 

Workshops are also represented in the form of an extended
book~\cite{playfulbook}, translated into the languages of the project
consortium members and available for free download from the web platform.
Through the process of creating, testing, and refining the workshop
descriptions, it has been possible to acquire significant experience and
develop new ideas about computing, teaching, curriculum, classroom experiences
and applied pedagogy.  It is possible for someone to just pick up one of the
packaged workshops and implement it immediately, independently from the others
without looking at any of the surrounding materials.  However, the book
enhances this with links between the workshops, and relationships between
themes which have emerged in the workshop.  The book also contains sections on
the learning environment, and also feedback (formative and summative) and
assessment~\cite{black2010} tools. 

\section{Outcome and results}

During the two-year project, it has  been possible to accomplish both tangible and intangible results, briefly:

\begin{itemize}
\item Building a catalog of inspiring playful coding activities that are integrated and available in the project web platform~\cite{playfulweb};
\item A teacher's guide edited and published in book format (physical and digital)~\cite{playfulbook} that has been translated into 6 languages of the project consortium members. This includes the playful coding activities, explanation and discussion of the underlying pedagogical methodology, technical advice, ideas, suggestions and challenges;
\item Establishing a process methodology for generating, testing, refining and managing the workshops and activities that potentially enable this kind of educational resources to live and grow over time. The methodology was developed by the consortium, and has been used during the creation and refinement of project activities.
\end{itemize}

More intangible outputs are represented by exchange of ideas and experiencing
good practices, while conducting a deep inter-cultural dialogue. Firstly
through the close collaboration and involvement of the core  project partners.
Secondly through the exchange diffusion of the refined and tested workshop
packages and the sharing of good practices with schools and entities external
to the project. The overall project impact currently includes more than 45
talks, seminars, training sessions for external educators, activities in more
than 80 schools, with a total reach of more than 600 teachers and 4000
school-aged participants across five different countries.

%% The transnational meetings and face to face training activities
%have facilitated this, but we have also shared ideas and resources through
%on-line collaboration and video.  Secondly, externally, by working not only
%with schools who belong to the consortium but also schools peripheral
%to the project. We have generated an exchange of good practice as well as an
%impact on them: there have been 45 talks, seminars, and training sessions
%communicating to educators outside of the project. We have also tested and run
%activities in more than 80 schools, reaching more than 600 teachers and 4000
%school-aged participants.

The analysis of the feedback survey collected from teachers participating in
the project shows that the playful approach was positively valued by our
collaborators.  The most common positive themes mentioned in the feedback, both
from written questionnaires and from focus group meetings, was that the
open-ended nature of creative workshops allow the whole class to contribute to
the playful activity.  Feedback analysis highlighted general positive impacts
upon students, teachers and schools. Negative comments were rare. The only
negative theme to remark upon was the teacher self-confidence bias: before
seeing the activities in action, several teachers were wary of trying to
implement them in class. Almost univesally, they later revise their
opinion.

One example of feedback summarizing the project is the following quote from a
Romanian high-school teacher: 

\emph{``Children perceived the whole activity more as a game and 
less as a typical school activity.}'' \ldots ``\emph{When we talked 
about Scratch, I told them 
it's meant to make Math or other school subjects easy to learn. 
I got their attention when I told them they can even draw or 
ride a bike, or fly a kite, whatever they want. The sky is the limit. 
In the end the students were amazed''.} 

The feedback from universities was also strong. Academics from all five
participating countries have visited local schools to share these activities.
These schools have used the workshops, at first with assistance from university
and the project team, but in many cases they have gone on to adopt the
activities independently regardless of country of origin. For instance, many
months after the official end of the project, British schools still regularly
run activities that were originated in Romania, Italy or Spain, and vice versa.
Indeed the ``Poetry animation with Scratch'' workshop, which originated in
Romania, has been run as an annual Welsh schools national contest, with tens of schools
across Wales taking part 2017-2019.

In summary, the robustness of the  design, testing and refining process has
enabled activities from across the project to be implemented successfully in
Romania, UK, France, Italy and Spain. When asking teachers and pupils about
what they have learned, the kind of answers collected have a clear common theme
-- the use of play and creativity has unlocked the potential of computational
thinking and coding across the curriculum.

\section{Conclusions}

In this paper a framework for transnational collaboration on teaching resources
has been presented, that has enabled the production of a set of workshops
encouraging computational thinking, and a guidebook for engaging schools to
reproduce, adapt, share and contribute to extend the proposed activities.
Ideas from computational thinking and learning through play have been
integrated in a computational context.  The three-phase workshop testing
process approach (paper-based review, practical activity review in the
classroom, and then face-to-face observed activity) has allowed the development
and the delivery of high quality workshop activities with real robustness to
situational variation.  The process developed has enabled a very productive
collaboration and resource sharing amongst a large number of transnational
partners and external participants, despite different computational, cultural,
economic and financial contexts.

Looking forward and beyond the project end: the  teachers' guide book has been
downloaded more than 4000 times across all six languages. Feedback on the book
is also continuously collected.  Continuing activity on the playful coding
platform represents the final, tangible output of the  two-year project, which
will be improved and extended.  The project partners are still contributing
with designing and iteratively improving new activities, and by inviting new
schools and teachers to join in testing and submitting activities for schools
to use.  This represents the foundation of an effective living library of
methods and tools for engaging young minds with creative computing. 

% I WOULD END the article  here
%Throughout this project participants has been asked to do different things,
%promoted complementary activities, encouraged children to learn from each
%other, to discover by themselves, to work in teams, and in the end the project
%promoted deep and playful learning.  \end{document}  % This is where a 'short'
%article might terminate

%ACKNOWLEDGMENTS are optional
\section{Acknowledgments}

We would like to thank all of the academics and teachers who directly supported
this project by writing, testing, and feeding into our activities and their
evaluation: Abir Zanzouri, Alexandrine Beya, Amanda Clare, Anna Ferrarons, Anna
Rhys Davies, Anna Roura Rabionet, Antoanela Dan, Caterina Lombardi,
Cioaca Norica, Dan Selisteanu, David Strubel, Delia Pirvu, Dorel Lupu, Elvira
Popescu, Erin Good, Esther Villarroya, Eugen Ganea, Ferran Jambert, Guillaume
Lemaitre, Gwenan Philips, Joan Massich, Laura Coravu, Mariona Niell, Mark Neal,
Marta Peracaula, Martina Sabatini, Martin Nelmes, Meritxell Estebanell, Mireia
Frigola, Mireia Pomar, Mohamed Belkacemi, Nigel Hardy, Ouadi
Beya, Samantha Roberts, Tegid Owen, Tomi Rowlands and Wayne Aubrey. We would
also like to thank all of the teachers and school children who tested and
evaluated our workshops. This work was supported by \textbf{EU Erasmus+
2014-1-ES01-KA201-004462}

\bibliographystyle{IEEEtran}
\bibliography{sigproc}


%\vfill


\end{document}